\documentclass[a4paper, amsfonts, amssymb, amsmath, reprint, showkeys, nofootinbib, twoside]{revtex4-1}
\usepackage[english]{babel}
\usepackage[utf8]{inputenc}
\usepackage[colorinlistoftodos, color=green!40, prependcaption]{todonotes}
\usepackage[pdftex, pdftitle={Article}, pdfauthor={Author}]{hyperref} 
\begin{document}
\title{Blockade of vortex flow by thermal fluctuations in atomically thin clean-limit superconductors}

\author{Avishai Benyamini$^{1}$ }
\email[Correspondence email address: ]{avishaiben@gmail.com}
\author{Dante M. Kennes$^{2}$ }
\author{Evan Telford$^{3}$ }
\author{Kenji Watanabe$^{4}$} 
\author{Takashi Taniguchi$^{4}$} 
\author{Andrew Millis$^{3}$}
\author{James Hone$^{1}$}
\author{Cory R. Dean$^{3}$}
\author{Abhay Pasupathy$^{3}$}

\affiliation{$^{1}$Department of Mechanical Engineering, Columbia University, New York, NY, USA}
\affiliation{$^{2}$Technische Universit\"at Braunschweig, Institut f\"ur Mathematische Physik, Germany}
\affiliation{$^{3}$Department of Physics, Columbia University, New York, NY, USA}
\affiliation{$^{4}$National Institute for Materials Science, 1-1 Namiki, Tsukuba 305-0044, Japan}

\date{\today} 

\begin{abstract}
Resistance in superconductors arises from the motion of vortices driven by flowing supercurrents or external electromagnetic fields and may be strongly affected by thermal or quantum fluctuations. The common expectation borne out in previous experiments  is that as the temperature is lowered, vortex motion is suppressed, leading to a decreased resistance. A new generation of materials provides access to the previously inaccessible regime of clean-limit superconductivity in atomically thin superconducting layers. We show experimentally that for few-layer 2H-NbSe$_2$ the resistance below the superconducting transition temperature may be non-monotonic, passing through a minimum and then increasing again as temperature is decreased further. The effects exists over a wide range of current and magnetic fields, but is most pronounced in monolayer devices at intermediate currents. Analytical and numerical calculations confirm that the findings can be understood in a two-fluid vortex model, in which a fraction of vortices flow in channels while the rest are pinned but thermally fluctuating in position. We show theoretically that the pinned, fluctuating vortices effectively control the mobility of the free vortices. The findings provide a new perspective on fundamental questions of vortex mobility and dissipation in superconductors.
\end{abstract}


\maketitle

The transport properties of superconductors penetrated by magnetic fields are determined by the dynamics of vortices, which are topological defects in the superconducting order parameter. Vortices may be introduced by magnetic fields $B$ \cite{abrikosov1957} or thermal fluctuations \cite{tinkham2004introduction} and may be free to move or may be pinned to defect sites. If vortices are not present or are pinned, the material is superfluid and can carry a current without dissipation. Vortex motion leads to a non-vanishing DC resistance and has been extensively studied in the last half century \cite{blatter1994}; one aspect of current interest is that vortex motion creates dissipation and decoherence, fundamental to superconductor-based quantum information technology.  The motion of vortices is believed to be overdamped on transport timescales and  the key parameters describing the resistivity $\rho$ are the density of mobile vortices $n_{\rm mobile}$ and the vortex mobility $\mu$, so that $\rho=\Phi_0^2\;n_{\rm mobile}\;\mu$ (see section 1 in supplementary materials; note that this formula assumes that, as is the case for most superconductors, the vortex hall angle is negligible).  An extensive literature exists on mechanisms affecting vortex mobility  in different limits \cite{tinkham1964,bardeen1965,clem1968,chow1970,larkin1998}. The number of mobile vortices is affected by disorder, which pins vortices, interactions, which lead to collective pinning, and the applied current, which may if sufficiently strong detach some or all vortices from pinning sites \cite{strnad1964dissipative,tinkham2004introduction,blatter1994}. Particularly in thin films, pinned and mobile vortices may coexist \cite{jensen1988lattice,hellerqvist1996vortex,embon2017imaging,benyamini2019}. In superconductors studied previously, the vortex mobility is not strongly temperature dependent, and the density of mobile vortices remains constant or decreases as temperature is decreased, leading to a resistance decreasing monotonically with temperature. In this work, we identify a new regime of vortex physics occurring in ultra-thin, clean limit superconductors exhibiting a previously unreported regime of vortex motion characterized by a nonmonotonic temperature dependence of the resistivity with an increase in resistance as temperature goes to zero, $T\rightarrow 0$,  and a corresponding anomalous enhancement in dissipation.

\begin{figure*}
\includegraphics{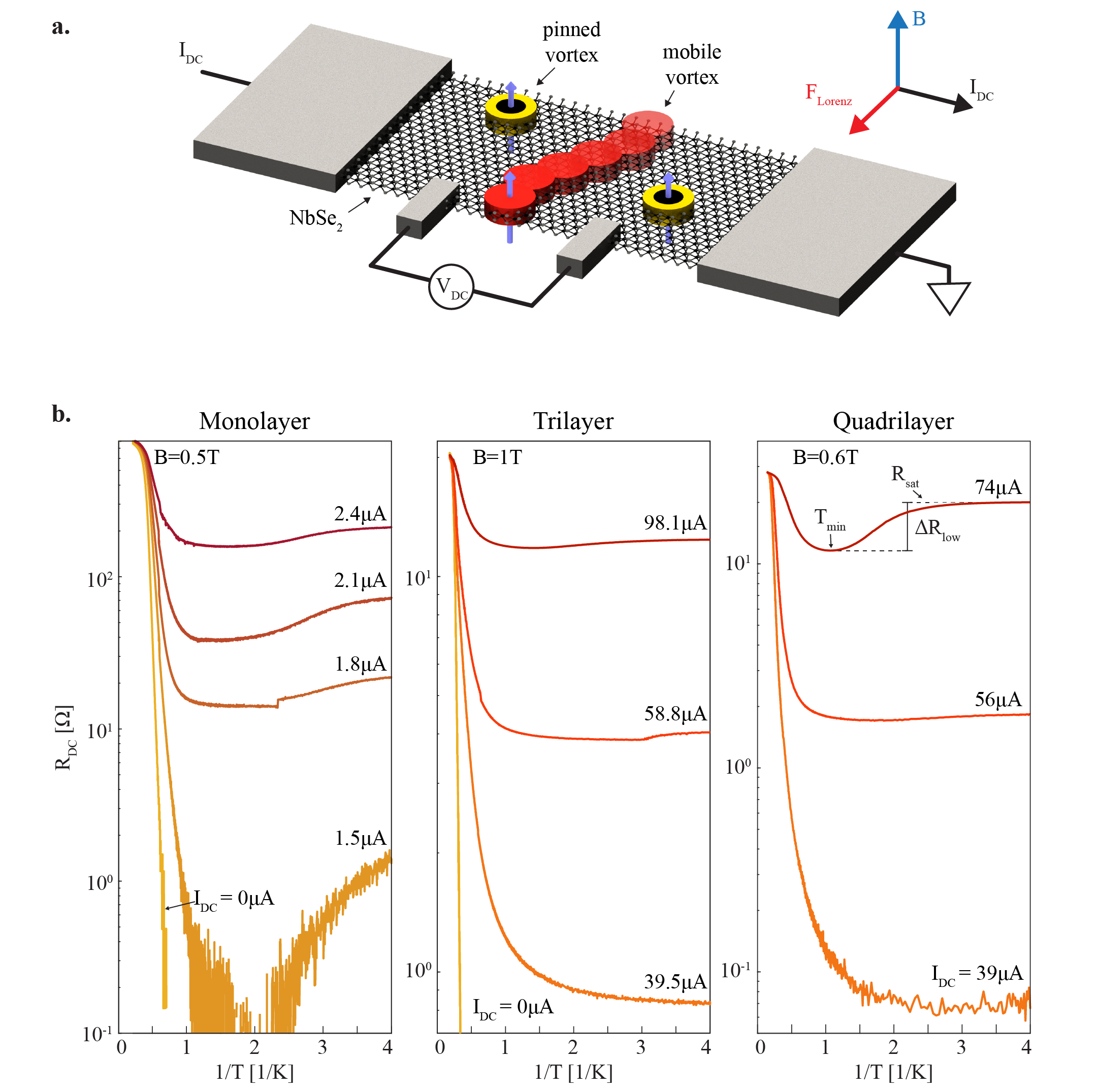}
\caption{Increase of resistance at low temperature. (a) Illustration of the device and measurement configuration. Pinned and mobile vortices are illustrated by yellow and red disks. The black dots in the middle of the pinned vortices are pinning sites. Blue arrows indicate the attached quantum of flux. The direction of the magnetic field $B$, the DC current and the Lorentz force are noted. (b) Line traces of resistance, $R_{\rm DC}$, as a function of inverse temperature at a finite magnetic field and different $I_{\rm DC}$ for monolayer, trilayer and quadrilayer devices. At $I_{\rm DC}=0$ the resistance is measured by a small AC excitation.}
\end{figure*}

We study ultra-thin samples of the layered superconductor 2H-NbSe$_2$~\cite{frindt1972,xi2015} using devices made using the 'via method' for lithography free contacts and preservation \cite{telford2018}. The via method enables the creation of low contact resistance and robust devices despite the air sensitivity of 2H-NbSe$_2$ flakes. We present here data from three devices, shown schematically in figure 1a, whose superconducting regions contain one, three or four atomic planes of  NbSe$_2$ respectively. All of our devices are in the clean limit, $\xi<l_{\rm mfp}$ ($\xi$ being the in-plane coherence length estimated from the $T=0$ limit of the critical magnetic field $H_{\rm c2}$ and $l_{\rm mfp}$ being the electron mean free path); see section 2 in supplementary information for additional details. 

We induce vortices by applying a magnetic field $B$; the extreme two-dimensionality of our samples means that the areal density of field-induced vortices is $n_v=B/\Phi_0$ where $\Phi_0=\frac{h}{2e}$ is the superconducting flux quantum. The vortex dissipative state is probed by applying a DC current, $I_{\rm DC}$. The current produces a Lorentz force of magnitude F$_{\rm Lorentz}=J \Phi_0 t$ (current density $J=I_{\rm DC}/wt$  where $w$ and $t$ are the sample width and thickness) on each vortex, causing some vortices to move.  We characterize the resulting dissipation by the measured resistance, $R_{\rm DC}=V_{\rm DC}/I_{\rm DC}$. In the resistance measurement, low temperature electronic filters with a cutoff frequency of $5$kHz were used to prevent measurement artifacts from electromagnetic noise \cite{tamir2019}. The measured devices are uniform and show the same behavior of $R_{\rm DC}(T;I_{\rm DC},B)$ measured between different contacts along the current direction, see figure S1, indicating that our results are intrinsic and not artifacts stemming from non-uniformity or hysteresis effects \cite{kwasnitza1990,petrov1992}.


Figure 1b shows the resistance in log-scale, plotted vs inverse temperature, for different measurement currents and at finite field for three devices with a one, three and four unit cell thick superconductors. For all samples the behavior is activated, $\log(R_{\rm DC})\propto-E_g/T$, in the $I_{\rm DC}\rightarrow 0$ regime measured by a small AC excitation, with the activation energy increasing linearly with the number of layers \cite{benyamini2019}. This behavior is expected: at very low currents the vortex lattice is collectively pinned and resistance arises from thermally activated defects in the pinned vortex lattice; the pinning energy scales with the thickness-integrated superfluid stiffness which increases rapidly with number of layers. For small but non-negligible applied currents, the resistivity remains monotonically decreasing, but saturates at a non-zero value $R_{\rm sat}$ corresponding to a flux flow regime  discussed elsewhere \cite{benyamini2019}. This behavior is clearly visible for the trilayer and quadrilayer devices at  $I_{\rm DC}\approx 39\mu A$. As the current is further increased, the resistance minimum develops. We see a sharp onset for thinner devices and smaller currents and a more continuous behavior for thicker devices and larger currents. We denote by $T_{\rm min}$ the temperature at which the resistance takes its minimum value and by $\Delta R_{\rm low}$ the increase of resistance from the minimum, to the low temperature saturated value, $R_{\rm sat}$. The definitions are shown in the right panel of figure 1b. As the current increases $\Delta R_{\rm low}$ first increases up to a maximal value and then decreases to zero. For sufficiently large currents (but still below the critical current) the resistivity again becomes monotonic. The observed phenomenology with increased $I_{\rm DC}$ is consistent across all magnetic fields, see figure S3, and is observed to be as large as 30\% of the normal state resistance, $R_{\rm N}$, for the measured devices (in the lower current regimes where a sharp increase of resistance is observed, a sharp transition is also observed at higher temperatures which we denote $\Delta R_{\rm high}$, see figure S2).

\begin{figure*}
\includegraphics{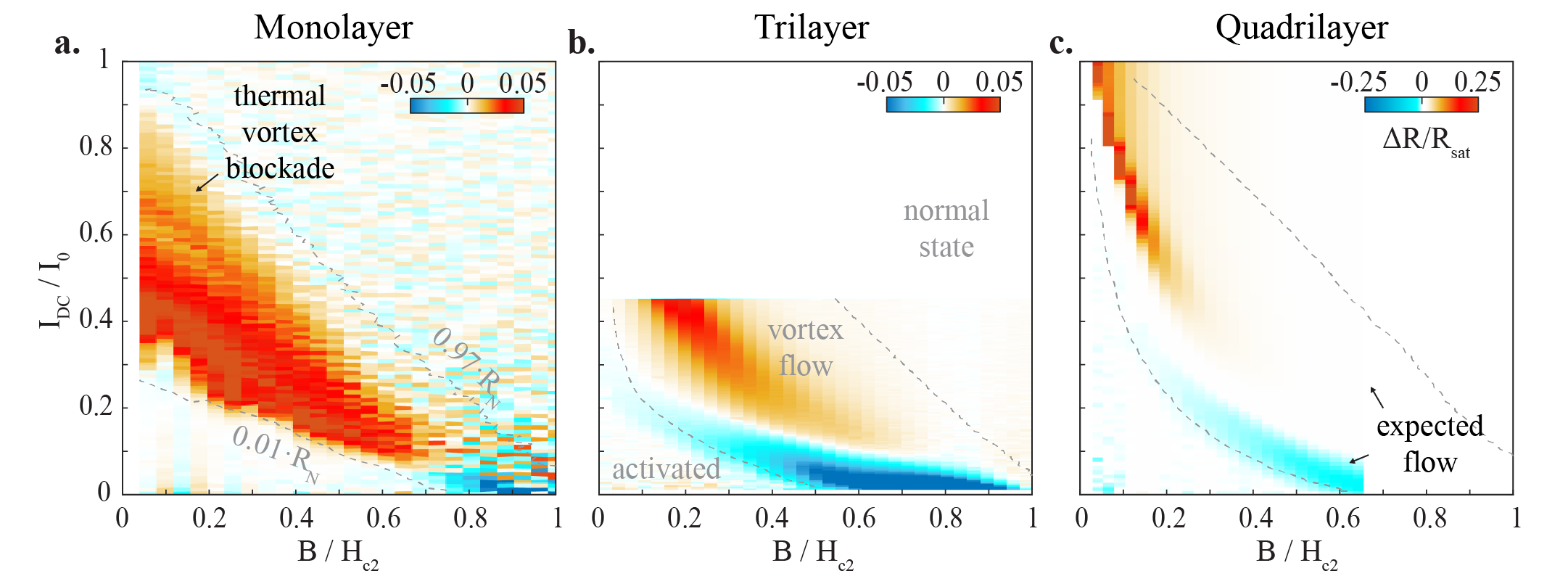}
\caption{Increase of resistance shifts to lower magnetic fields and higher currents with increased thickness. (a), (b) and (c) show qualitatively the steady-state phase diagrams of $\Delta R_{\rm low}/R_{\rm sat}$, with $T_{\rm min}\sim 550, 600$ and $1000mK$ for the monolayer, trilayer and quadrilayer, as a function of DC current and magnetic field for a monolayer, trilayer and quadrilayer devices. We expect the diagram to be negative (blue) or zero (white) due to the monotonic behavior of the activated and flux flow regimes. The non-trivial regime of increased resistance at low temperature is shown in red. It is observed that as a function of layer number the non-trivial regime shrinks to smaller magnetic fields and higher currents. Dashed gray lines denote, at finite $B$, the transitions from activated to vortex flow and to the normal state with increased DC current.}
\end{figure*}

To visualize the regime where the increase of resistance is found and how it changes with layer number we present in figure 2 a color map of the normalized change in resistance between our lowest temperature, $T\sim 250mK$, and a fixed temperature around $T_{\rm min}\sim 550, 600$ and $1000mK$ for the monolayer, trilayer and quadrilayer devices respectively. The expected behavior of the resistance is a decrease or saturation with decreased temperature shown by blue and white color in the figures, respectively. The nonmonotonic behavior is shown in red in the color maps and reflects that the resistance increases at low temperature. We observe that in a monolayer most of the vortex flow regime displays this unexpected behavior. As the layer number is increased the regime of increased resistance is shifted to lower magnetic fields and higher currents.

Our understanding of the measured behavior begins from the observation  that our samples are in the clean limit, therefore characterized by dilute (on the scale of the coherence length) scattering sites. We assume that some fraction of the scattering sites are strong pinning centers, which trap individual vortices.  For very weak drive currents, the logarithmic vortex-vortex interactions lead to collective pinning of the entire vortex lattice, so that the linear response resistivity is exponential in inverse temperature, corresponding to thermal activation of defects in the collectively pinned vortex lattice. As the drive current is increased, some of the vortices become detached from the pinned lattice and can move, leading to a flux-flow resistivity that remains non-zero even at lowest  temperature. The crucial new insight is that the motion of the detached vortices is affected by scattering from the pinned vortices; in particular, as temperature increases, the pinned vortices fluctuate more and more strongly about their pinned positions, leading to scattering that increases as temperature increases and hence lead both to a strongly temperature-dependent vortex mobility and to a decrease in the number of mobile vortices as the temperature is increased.

\begin{figure*}
\includegraphics{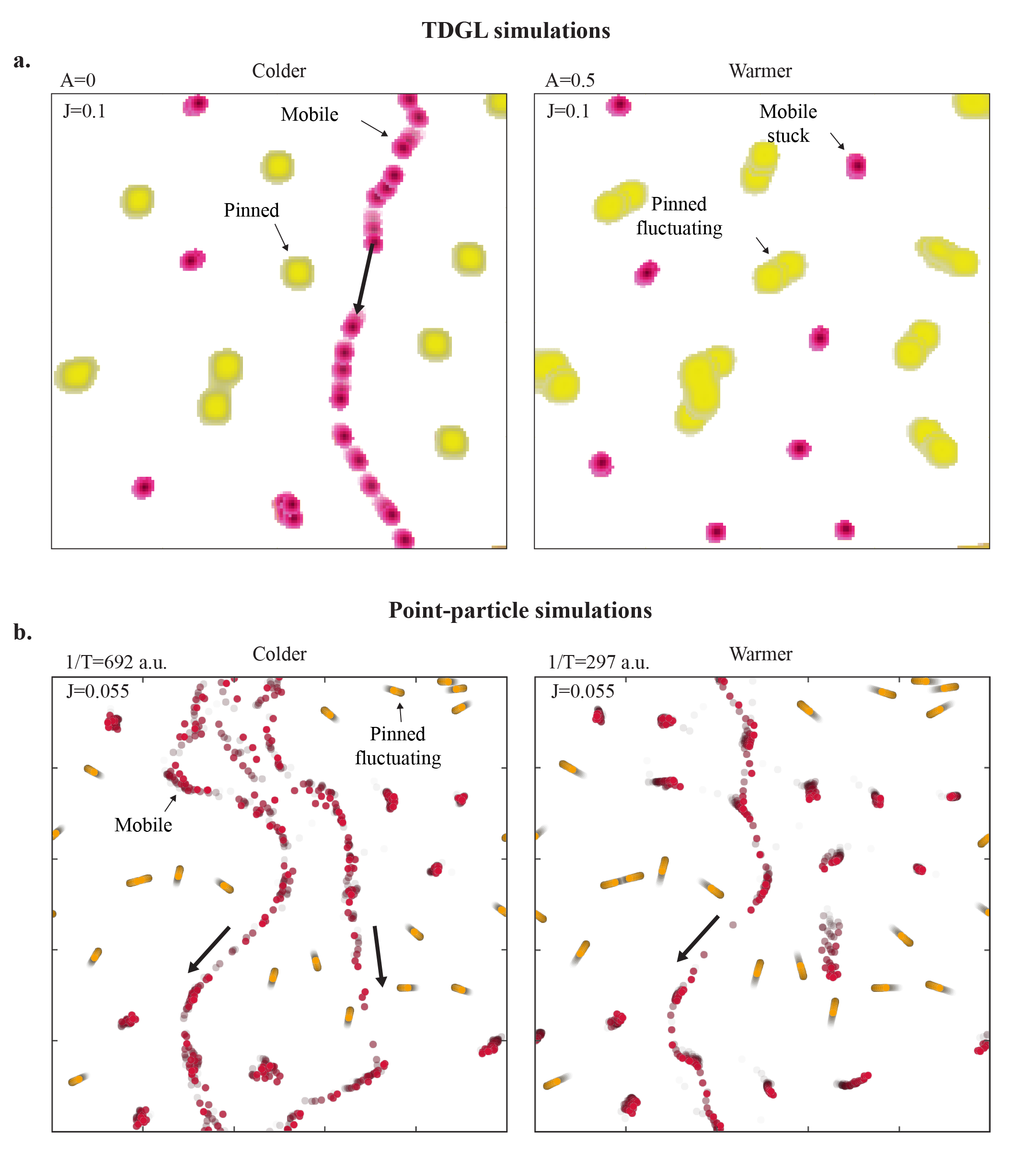}
\caption{Simulations of vortex dynamics in the presence of pinning and fluctuations. (a) Overlaid snapshots at different times from the TDGL simulation for $J=0.1$ at two oscillation amplitudes (effective temperatures), $A=0$ ($T=0$ limit) and $A=0.5$ (finite $T$). The vortices are false colored for visualization. The yellow vortices are the pinned fluctuating vortices and the cherry colored ones are the mobile vortices. A single vortex conducting channel is observed for the 'colder' scenario, while at 'warmer' temperature all vortices are immobile due to the oscillations of the yellow vortices at the pinning sites. The online movies, M1 and M2, show the superconducting gap with red for full gap and blue for zero gap (see supplemental information). (b) Similar to panel a for the point-particle simulations. Free and pinned particles are shown by red and yellow dots, respectively. Arrows indicate vortex motion direction. The full movies are available on-line as movies M3 and M4.}
\end{figure*}

To investigate this scenario in detail we have numerically solved the time-dependent-Ginzburg-Landau (TDGL) equations for a two dimensional strip geometry including a magnetic field, which sets the number of vortices,  and strong pinning centers, which may trap some of the vortices; see section 3 in supplementary materials for details.  The effect of thermal fluctuations is simulated by requiring that  pinned vortices oscillate  with an amplitude  that increases with temperature, $T\sim\Omega^2A^2$ ($\Omega$ and $A$ are the frequency and amplitude correspondingly). From our simulations we compute the resistance (voltage drop across the computational sample divided by imposed current) and generate movies of the time and space dependent magnitude of the order parameter $\Delta$. In the movies  vortex positions are clearly revealed as  the local suppressions of $\Delta$. The simulations capture the same qualitative dependence observed in the experiment where as the oscillations amplitude is increased the average velocity of the mobile vortices decreases and vortices may possibly even stop moving altogether, see figure 3a and supplementary movies M1 and M2.  The TDGL computations are expensive. We found that the qualitative behavior of the vortices found in our TDGL simulations is well represented  by  a simplified model in which the vortices are idealized as point particles interacting by the logarithmic potential found in the TDGL equations, see section 4 in supplementary materials. We split the particles into two types, mobile and pinned, and analyze again what happens to the mobile particles if the pinned particles oscillate. We quantify the experimentally measured resistance by the average rate that free particles leave the sample divided by the drive force. We have used this computationally more tractable model  to access a wider range of temperatures, current strengths and sample sizes.  

The key feature of both the TDGL and the particle simulations is that at intermediate current strengths the mobile vortices are confined to channels confined by regions of bound vortices, as also found in prior work \cite{hellerqvist1996vortex,embon2017imaging}.  The simulations (e.g. the TDGL simulations shown in the lower panel  of figure 3b  and in supplementary movies M3 and M4) show that as temperature increases the mobility in each channel continuously decreases until the channel abruptly stops conducting vortices. In figure 3b at the colder temperature two paths are observed, while the warmer panel shows a single path as the second was shut off due to the thermal motion of the pinned particles.   Figure 4a shows simulations for fixed numbers of free and bound particles at two different drive currents.  The simulation shows the same behavior as the experiment (overlaid in grey).  In particular, at weaker drive current there is only one channel and a discontinuous transition is observed as the channel is closed by increasing fluctuations as temperature is increased, whereas at stronger drive currents  additional channels are added, which may  interconnect, leading to a more complicated network structure that  gives a smoother resistive transition as observed  in the experiment and theory. 

\begin{figure*}
\includegraphics{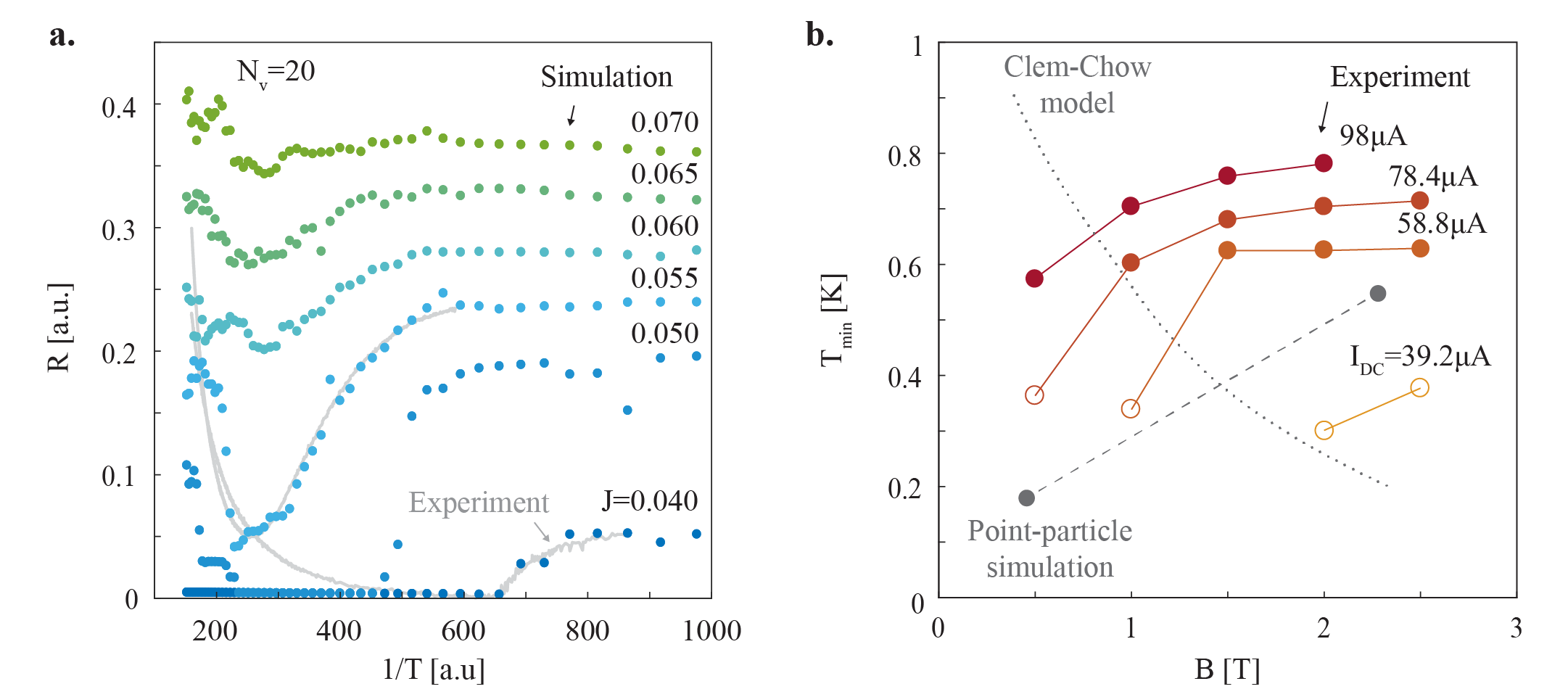}
\caption{Blockade of vortex motion due to thermal fluctuations. (a) Calculated resistance obtained from the simulation as a function of inverse temperature for different currents. Similar trends to those in the experiment (overlayed in gray) are observed. (b) $T_{\rm min}$ dependence on magnetic field for difference DC currents. Open and full circles denote discontinuous and continuous increase of resistance at $T_{\rm min}$ respectively. The dotted gray line represent the theoretical curve extracted from Clem-Chow theory~\cite{chow1970} with different x,y units and limits. The x and y-axis are in units of $B/H_{\rm c2}$ and $T/T_{\rm c}$ with limits (0,0.35) and (0.35,0.5) respectively. The gray points and dashed line are from the point-particle simulations. The x and y-axis are in the same arbitrary units as in panel a for $N_{\rm V}$ and $T$ with limits (17.5,20.5) and (0,0.01), respectively.}
\end{figure*}

An alternative explanation for a resistance which increases as temperature is decreased was provided by Clem and Chow \cite{clem1968,chow1970}. The Clem-Chow model assumes the dirty superconductor limit at which temperature gradients can be defined on the scale of a vortex core and may not hold for our devices which are in the clean limit.  A key difference between the Clem-Chow model and our results is the magnetic field dependence of the temperature of the resistance minimum $T_{\rm min}$. As shown in figure 4b in both experiment and the proposed thermal-scattering model $T_{\rm min}$ increases with increased $B$, see figure S4, while in Clem-Chow model $T_{\rm min}$ decreases. The thermal-scattering model reproduces additional similarities as evident by the comparison in figure 4a and as was discussed above. Additionally $T_{\rm min}$ and $R_{\rm sat}$ are observed in both experiment and the theoretical model to shift to higher temperature with increased $J$ as shown in figure 1b,c and figure S3, respectively.

The data exhibit a strong dependence on the number of superconducting layers in the device: as the number of layers is increased, the non-monotonic region is confined to lower vortex densities and stronger Lorentz force. We understand this as a consequence of two related effects. First, the c-axis coherence length is longer than four unit cells, so a vortex is an essentially vertical column and can be pinned by a pinning site in any layer. Thus the effective density of pinning sites is increased, meaning that a stronger current is required to displace a vortex. Further, the vortex-vortex interaction also increases linearly with the layer number   ~\cite{benyamini2019}; this increases the collective pinning and inhibits formation of channels at higher vortex densities.  These data suggest that the key physics required to observe the non-monotonic temperature dependence of dissipation includes two dimensionality, low superfluid stiffness (hence weak intervortex interactions), and clean limit (so dilute strong-pinning centers).


Understanding vortex dissipation mechanisms is a crucial step to manifesting and controlling emergent phenomena in superconductors and their interfaces with other materials. A vortex is a mesoscopic system with a natural normal-superconductor interface. Such interfaces can lead to novel electronic phenomena such as  Majorana fermions \cite{majorana1937} when the superconductor is topological \cite{fu2008,sun2017,Claassen2019}. In the clean limit, proximity of the superconductor to the normal cores may also lead to new emergent phenomena in vortices~\cite{choi1994,stern1994,rozhkov1996,onogi1996}. Dissipative vortex motion will hinder exploration of such states and their usage for quantum information applications. Traditionally, improving materials quality by removing disorder is the route to improving quantum coherence properties. Our work shows that in crystalline 2D superconductors, the relative weakness of disorder plays the opposite role by enabling free motion of vortices and thus enhancing dissipation. In future applications of clean 2D superconductors \cite{saito2017}, it is therefore important to engineer disorder that is effective at eliminating vortex dissipation, while at the same time not introducing other degrees of freedom that can couple to the superconductor and cause decoherence. This remains an open problem. 

\section*{Acknowledgments}
This work was primarily supported by the NSF MRSEC program through Columbia in the Center for Precision Assembly of Superstratic and Superatomic Solids (DMR-1420634) and by Honda Research Institute USA Inc. DMK was supported by the Deutsche Forschungsgemeinschaft through the Emmy Noether program (KA 3360/2-1). Simulations were performed on Columbia University's Habanero Computing Cluster. AJM performed part of the work on this paper at the Aspen Center for Physics, which is supported by National Science Foundation grant PHY-1607611.

\bibliography{refs}

\appendix*

\end{document}



\baselineskip24pt


\maketitle 


\section*{\large This PDF file includes:}

Supplementary information sections 1-4\\
Figs. S1 to S4\\

\section*{\large Other Supplementary Materials for this manuscript include the following:}
Movies M1 to M4

\clearpage

\renewcommand{\thesection}{}
\renewcommand{\thesubsection}{\arabic{subsection}}


\subsection{Linear V-I relation of mobile vortices}
Each vortex carries a quantum of flux which generates a winding of the superconducting phase around the vortex core. The motion of vortices changes the superconducting phase in space and time generating an electrostatic potential given by the Josephson relation, $V=\frac{\hbar}{2e}\frac{d}{dt}\Delta\varphi$ ($\hbar$, $e$ and $\varphi$ are the reduced Planck constant, the electron charge and the superconductor phase respectively). The change in the phase over time can be written in terms of the mobile vortex density, $n_{\rm mobile}$, and the mean vortex velocity, $v_y$, giving an electrostatic potential $V=L\Phi_0n_{\rm mobile}v_y$ ($L$ the distance between voltage probes and $\Phi_0$ the quantum of flux). The overdamped vortex velocity is related to the Lorentz force by the vortex mobility, $\mu$, $v_y=\mu F_L$, and to the current through $F_L=J\Phi_0t$ ($t$ the film thickness, $J=I/wt$, $I$ the sourced current and $w$ the sample width). Overall giving a linear voltage-current relation that can be written in terms of a resistance, $R_{\rm mobile}=\frac{L}{w}\Phi_0^2\;n_{\rm mobile}\;\mu$.

\subsection{Parameters of measured devices}
In the table below we summarize parameters of the measured devices: the layer number, the 2D mean free path $l_{\rm 2D}=\frac{h}{e^2}\frac{1}{\sqrt{2\pi n_1}R_1}$ ($n_1$ and $R_1$ are the electron density and resistance per layer), and the coherence length estimated from the $T\rightarrow0$ limit of $H_{c2}(0)$, $\xi^2=\frac{\Phi_0}{2\pi H_{c2}(0)}$.
\begin{center}
\begin{tabular}{ |c|c|c|c|c|c| } 
\hline
Layer No. & T$_{\rm c}$[K] & R[$\Omega$] & $l_{\rm 2D} [nm]$ & H$_{\rm c2}$[T] & $\xi [nm]$ \\
\hline
1 & 3.6 & 206 & 16 & 4 & 9 \\
3 & 5.5 & 21 & 55 & 4 & 9 \\ 
4 & 6.2 & 28 & 31 & 5.3 & 7.5 \\ 
\hline
\end{tabular}
\end{center}

\subsection{Time-Dependent-Ginzburg-Landau simulations}

To model the dynamics of driven vortices we employ a deterministic set of TDGL equations to describe the 2D superconductor in the presence of a perpendicular magnetic field and a sourced current \cite{ivlev84,kennes17}.
We compute the complex superconducting order parameter $\Delta=\left|\Delta\right| e^{i\phi}$, the charge density $\rho$ as well as the current density $\vec{j}$ in dependence of time and space in the 2D sample. The electromagnetic fields are represented by the vector potential $\vec{A}(t)$ as well as the scalar potential $\Theta(t)$.  We set units by choosing $\hbar=c=e=1$ such that the superconducting flux quantum $\Phi_0\equiv hc/2e=\pi$. 

The equations ({\it 34-36\/}) are thus
\begin{align}
&\frac{1}{ D}\left(\partial_t+2i\Psi\right)\Delta=\frac{1}{\xi^{2}\beta}\Delta\left[r-\beta |\Delta|^2\right]+\left[\vec{\nabla} -  2i\vec{A}(t)\right]^2 \Delta
\label{eq:TDGL}
\\
&\rho=\frac{\Psi-\Theta}{4\pi\lambda_{\rm TF}^2}
\label{eq:TDGLrho}
\\
&j=\sigma\left(-\nabla \Psi-\partial_t A(t)\right)+\frac{\sigma}{\tau_s}{\rm Re}\left[\Delta^*\left(\frac{\nabla}{i}-2A\right)\Delta\right] 
\label{eq:TDGLj}
\end{align}
Here $D$ is the normal state diffusion constant, $\Psi$ is the electrochemical potential per electron charge,  $\xi=\sqrt{6 D\tau_s}$  is related to the superconducting coherence length $\xi_0=\xi/\sqrt{r/\beta}$, where $\tau_s$ is the spin-flip scattering time, $\lambda_{\rm TF}$ is the Thomas-Fermi static charge screening length and  $\beta$ is a system dependent constant that sets the magnitude of the order parameter. For definiteness we  measure lengths in units of $\xi$ and time in units of $\xi^2/D$ (which we write simply as $D^{-1}$ since $\xi$ is our unit of length)  and choose parameters $\beta=1$, $\sigma=0.1 $, $\tau_sD=\frac{1}{6}$  and $\lambda_{\rm TF}^2/\xi^2=0.1$. We verified that choosing  other parameters does not affect the general conclusions. The equations are furthermore supplemented by the continuity equation  
\begin{equation}
\partial_t \rho+\vec{\nabla}\cdot\vec{j} =0
\label{eq:continuity}
\end{equation}
and the Poisson equation for the scalar potential 
\begin{equation}
\nabla^2\Theta=-4\pi\rho.
\label{eq:Poisson}
\end{equation}
Using  a finite elements approach we solve the coupled partial differential equations Eqs.~\eqref{eq:TDGL}-\eqref{eq:Poisson} with periodic boundary conditions in the y-direction while being open in the x-direction (choosing vanishing first derivatives at that boundary as a boundary condition). We thus concentrate on a torus geometry. We discretized time in steps of $D\Delta t =0.0001$, which we verified to be numerically  converged. For finite sourced current we choose the boundary conditions of the current density such that the one end of the open boundary acts as a particle source while the other acts as a drain $j_x(x=0,y,t)=j_x(x=L_x,y,t)=j_0$  with $L_x$ the size of the system in x-direction ($L_y$ denotes the size of the system in y-direction). The initial conditions for all other variables but the order parameter $\Delta$ are chosen to be zero at $t=0$ and drawn  from a random uniform distribution in the interval $[0,0.001]$ for $\Delta(x,y,t=0)$. Assuming Coulomb gauge we determine  $A$ from $\nabla\times A=B$ for given external magnetic field $B$. 

To model disorder pinning vorties we choose to allow for a position dependent $r(x,y)=r_0-\delta r(x,y)$, where $r_0=1$ sets the superfluid stiffness without disorder and $\delta r(x,y)$ describes the disorder effects. We use
\begin{equation}
\delta(x,y)=\sum\limits_{i=1}^N \delta_i \Theta\left(-|x-x_i|/\zeta-|y-y_i|/\zeta\right),
\end{equation}
where $N$ describes the total number of defects, $(x_i,y_i)$ denotes the position of the $i$-th defect and $\delta_i$ is the i-th defect's strength. 
We choose $N=10$, $\delta_i=5$ and draw $x_i$ and $y_i$ from a uniform distribution $(0,L_x]$ and  $(0,L_y]$, respectively. 

These randomly distributed defects will tend to pin some of the vortices in place. The TDGL equations are a mean-field theory so pinned vortices are simply frozen at the defect position. Therefore, to model the thermally induced motion of the vorteces around the defect position we move the defect position in a random direction by replacing $x_i\to x_i(t)=x_i+A n_i^x\sin(\Omega t) $ and  $y_i\to y_i(t)=y_i+A n_i^y\sin(\Omega t)$ with a randomly drawn direction vector $n_i=(n_i^x,n_i^y)$ with unit length for each pinning site. We choose $\Omega=0.05$ in our units of time. The temperature inducing thermal fluctuations of the pinend vortices is then to be identified via the kinetic energy $T\sim A^2\Omega^2$ and we study its effects by scanning $A$. We find that thermally activating the pinned vortices can hinder the vortex flow which is shown in figure 3 and in the corresponding movies M1 and M2.

\subsection{Modelling vortices as point like particles in a box}

As a simplification we treat the vorteces as point-like particle in a
box with periodic boundary conditions which we found to reproduce the main features of vortex motion as described by the full TDGL simulations. The vortices are separated into $N_{\rm mobile}$ moving and $N_{\rm pin}$ pinned ones. The pinned vortices follow the instantaneous positions of the strong pinning sites, where the oscillations mimic thermal activation. The position $R_{{\rm pin},i}(t)$ of the i-th pinned vortex thus follows  
\begin{equation}
R_{{\rm pin},i}(t)=\left(x_i+A n_i^x\sin(\Omega t),y_i+A n_i^y\sin(\Omega t)\right)
\end{equation}
with $x_i$ and $y_i$ drawn from a uniform distribution $(0,L_x]$ and  $(0,L_y]$, respectively as well as a 
randomly drawn direction vector $n_i=(n_i^x,n_i^y)$ with unit length as above.

The mobile vortices then follow over-damped dynamics with 
\begin{equation}
\frac{dR_{{\rm mobile},i}(t)}{dt}=\left(F_{J}+\sum_{j\neq i}F(R_{{\rm mobile},i},R_{j})\right),
\end{equation}
where $F_{J}$ is the forced exerted on the vortices due to the sourced current and the second term on the right hand side describes the Force exerted by the vortex $j$ (be it pinned or mobile) onto the vortex $i$, which is long-ranged $F(R_{{\rm mobile},i},R_{j})=C\frac{R_{{\rm mobile}, i}-R_j}{|R_{{\rm mobile}, i}-R_j|^2}$ with a characteristic $1/{|R_{{\rm mobile}, i}-R_j|}$ dependence. For definiteness we choose $F_J=(0,J)$, $\Omega=5$ and $C=0.1$.

\bibliography{refs}

\clearpage

\renewcommand{\figurename}{Extended Data Figure}

\begin{figure}
\includegraphics[width=1\linewidth]{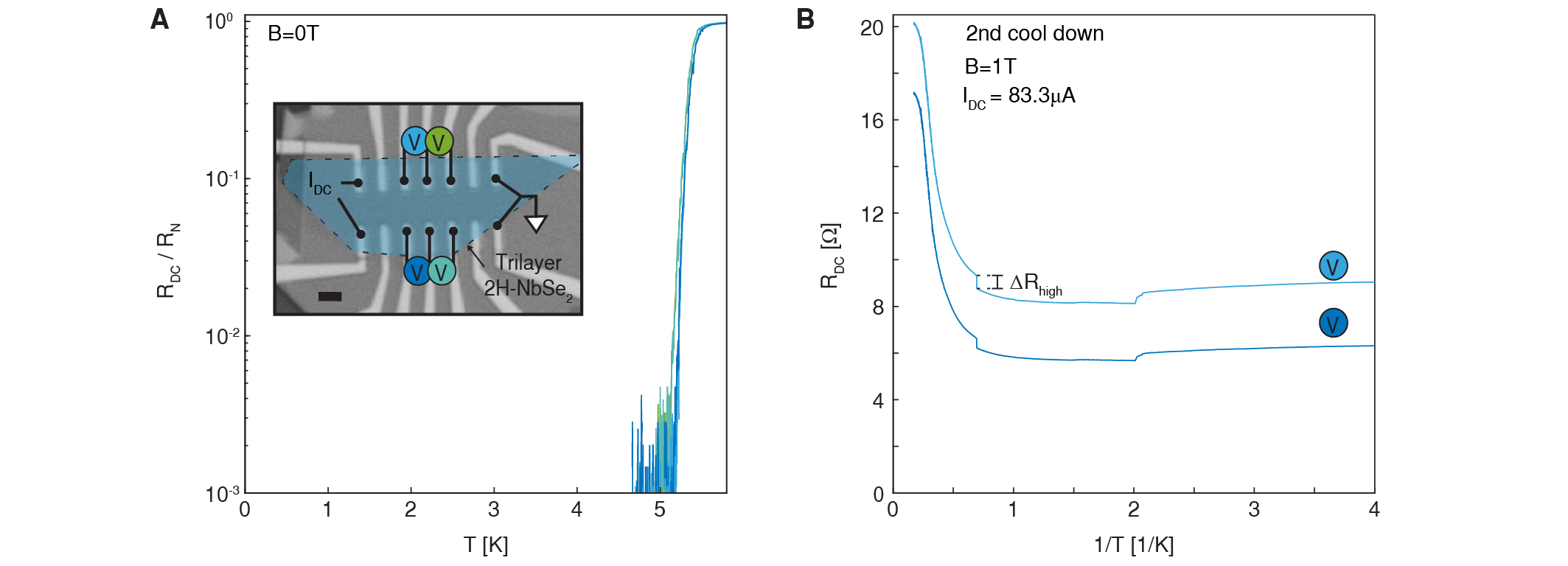} 
\noindent {\bf Figure S1: Device uniformity.} {\bf a.} Temperature dependence of the resistance at B=0 for various contacts around the measurement path. Parallel contacts are especially uniform while a small difference is observed between series contacts. The line color denote the contact as illustrated at the device picture. {\bf b.} Robustness of the increased resistance along the device. Data is shown for two neighboring contacts parallel to the current path. Line color uses the same scheme shown in panel a. The data is from a 2nd cool-down where the shape of the curve changed quantitatively but not qualitatively as shown in figure 1. 
\end{figure}

\begin{figure}
\includegraphics[width=1\linewidth]{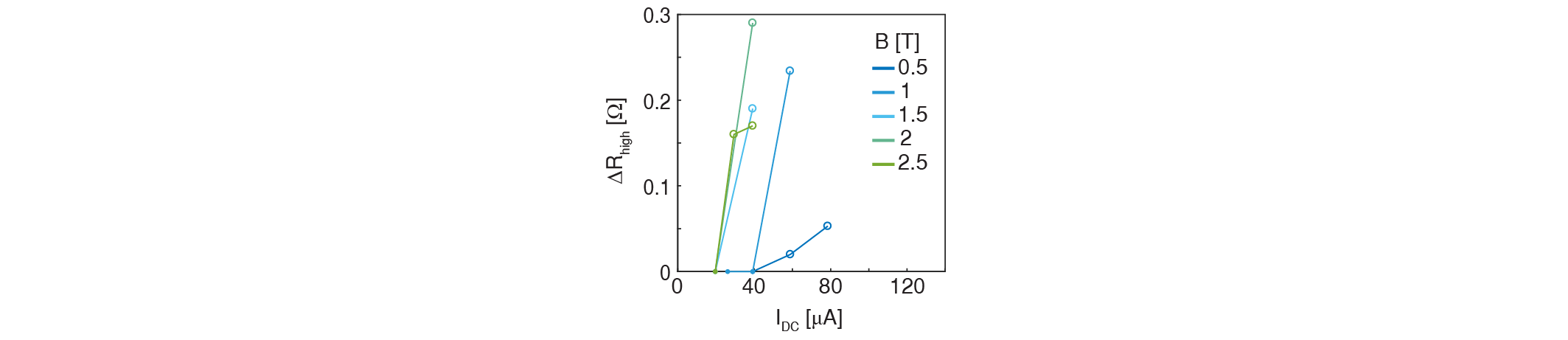} 
\noindent {\bf Figure S2: Current and magnetic field dependence of $\Delta R_{high}$ for trilayer device.} $\Delta R_{high}$ is defined as in figure S1b from the data in figure S3. Data points are shown where a sharp jump could be quantified.
\end{figure}

\begin{figure}
\includegraphics[width=1\linewidth]{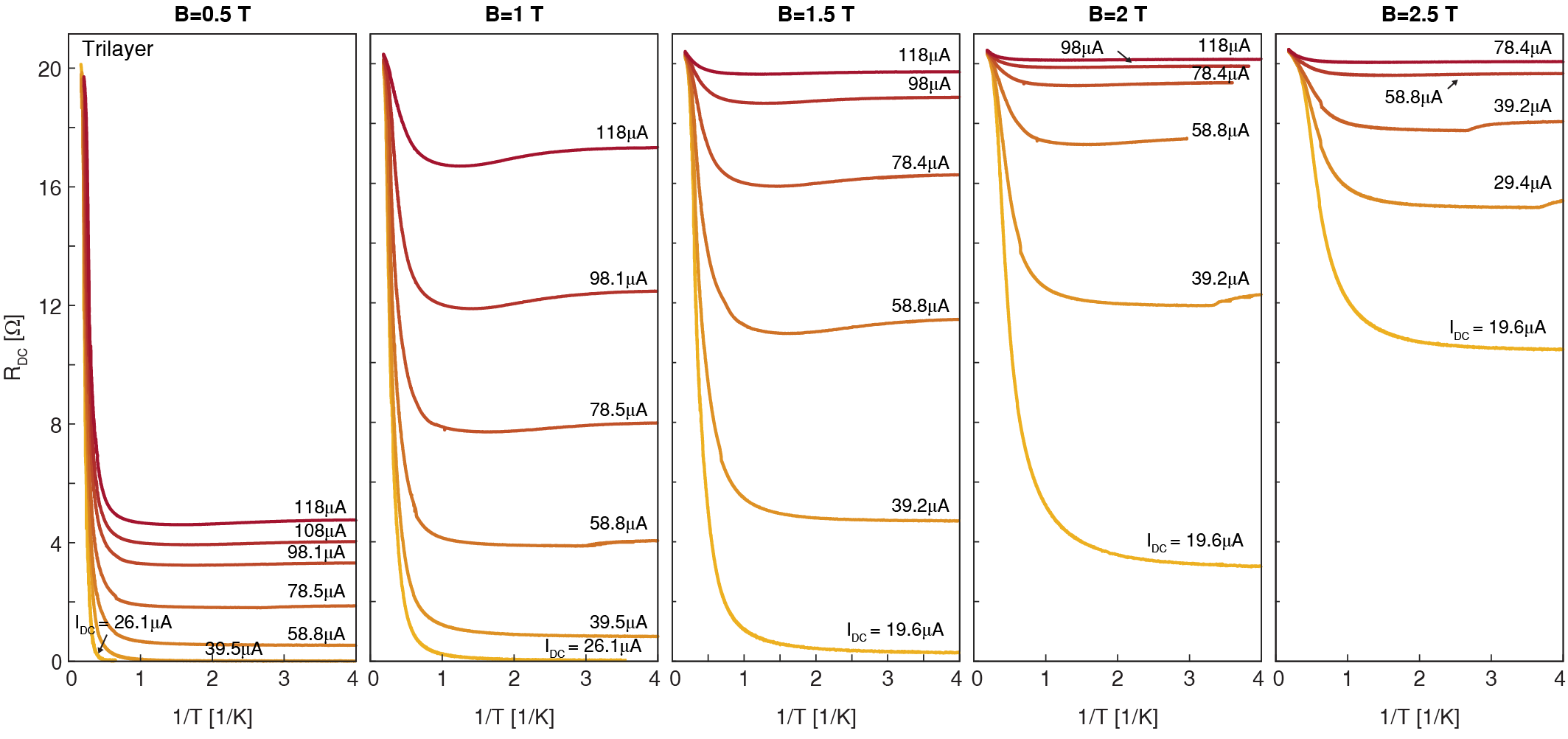} 
\noindent {\bf Figure S3: Temperature dependence of DC resistance at finite DC current for measured magnetic fields for trilayer device.}
\end{figure}

\begin{figure}
\includegraphics[width=1\linewidth]{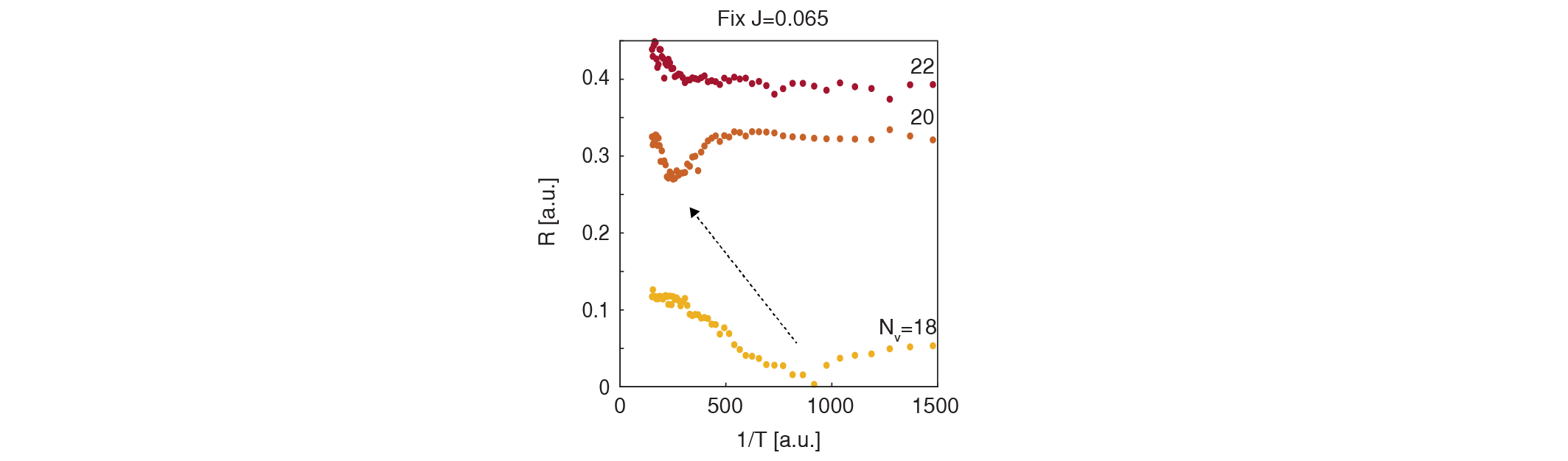} 
\noindent {\bf Figure S4: Shift in $T_{\rm min}$ with increased B in the simulation.} At fixed sourced current the minimum in temperature $T_{\rm min}$ shifts to larger values (smaller $1/T$). This is in accordance with the experimental finding of figure 4 of the main text.
\end{figure}